\documentclass[%
 aip,
 amsmath,amssymb,
 reprint,%
]{revtex4-1}

\usepackage{graphicx}
\usepackage{dcolumn}
\usepackage{bm}
\usepackage{amsmath}
\usepackage{mathrsfs}
\usepackage{amsfonts}
\usepackage{amssymb}
\usepackage{graphicx}
\usepackage{subfigure}
\usepackage{amsthm}

\usepackage[utf8]{inputenc}
\usepackage[T1]{fontenc}
\usepackage{mathptmx}
\usepackage{etoolbox}

\makeatletter
\def\@email#1#2{%
 \endgroup
 \patchcmd{\titleblock@produce}
  {\frontmatter@RRAPformat}
  {\frontmatter@RRAPformat{\produce@RRAP{*#1\href{mailto:#2}{#2}}}\frontmatter@RRAPformat}
  {}{}
}%
\makeatother
\begin{document}

\preprint{AIP/123-QED}

\title[H-theorem do-conjecture]{H-theorem do-conjecture}
\author{M. S\"uzen}
 \homepage{mehmet.suzen@physics.org}

\date{\today}

\begin{abstract}
A pedagogical formulation of Loschmidt's paradox and H-theorem is presented with basic notation
on occupancy on discrete states without invoking velocity collision operators. A conjecture,
so called {\it H-theorem do-conjecture}, is formulated. Causal inference perspective on the
dynamical evolution of classical many-particle system is invoked. This perspectice introduce
a probabilistic view on the state of the system conditioning on the thermodyamic
ensemble, i.e., function of state-variables representing the ensemble.  A numerical simulation
of random walkers for deterministic diffusion demonstrate the causal effect of interventional
ensemble, showing a dynamical behaviour as a test of the proposed conjecture. Moreover, the
chosen game like dynamics provides an accessible practical example, named
{\it Ising-Conway Entropy Game}, in order to demonstrate increase in entropy over time,
as a toy system of statistical physics.
\end{abstract}

\maketitle

\section{Introduction}

Philosophical foundations of statistical mechanics is rich in terms of
both theoretical and practical underpinnings of correspondence between dynamics in
small scales to collections in thermodynamic limit \cite{sklars93,styer04}. In this direction,
the most foundational argument that limits how physical processes evolve
over time is the Boltzmann's entropy \cite{boltzmann72} within the context of the second law of
theormodynamics.

A puzzling paradox is noted by Loschmidt-Zermelo on Boltzmann's H-Theorem \cite{boltzmann72, loschmidt77, wu75}
that H-function expressing velocity collisions in kinetic theory can't generate a time-irreversible macroscopic
dynamics while foundational microscopic mechanics is time-reversible.

Development of statistical mechanics overlaps with statistical inference was
initially noted by Jaynes \cite{jaynes57a,jaynes57b, jaynes03}  along with the  probabilistic techniques
on the trajectory evolution \cite{suezen18}. Recently, the connections of statistical physics
and deep learning are discussed \cite{suezen22, suezen21}. Along these lines, statistical
inference connections, recent mathematical tools introduced in doing causal
inference \cite{pearl09} provides an opportunity to revisit the formulation of H-theorem
via Gibbs's ensemble theory \cite{gibbs}.

We restrict the discussion of H-function as Boltzmann's entropy with a toy
pedagogical tool, so called dynamical evolution on discrete states \cite{kittel04,gibbs, susskind}.
This allows us to provide pedagogical definitions without loss of generality in invoking
a {\it do-calculus} \cite{pearl09, pearl95} perspective addressing Loschmidt's paradox.

\section{Differentiating Boltzmann's entropy}

The concept of entropy is quite diverse, initially introduced by Carnot-Clasius \cite{clausius, callen}
for operational efficiency of heat engines, thermodynamic entropy. However,
Boltzmann's interpretation has connected the classical mechanics of many-particle
system to thermodynamics, further refined by Gibbs \cite{gibbs}. Noticibly the distinction of Boltzmann's entropy against
other type-of entropies exists. Such as information entropy of Shannon's \cite{shannon},
Bekenstein's interpretation of a surface-area of a black-hole \cite{bekenstein} and von-Neumann's
quantum mechanical entropy \cite{vonn}. Here, we adhere to Boltzmann's definition on physical systems via so called
{\it accessible micro-states} \cite{kittel04} with Shannon's information capacity formulation.

\subsection{Occupancy on discrete states: Lattice}

Lattice dynamics is one of the land mark tool in understanding classical multi-particle mechanics \cite{ising, susskind}
It is also used as a pedagogical tool in understanding statistical mechanics \cite{kittel04}. Counting accessible microstates
of a physical system is associated with entropy.

{\bf Definition 1}: An entropy of a macroscopic material is associated with larger number of states its constituent
elements take different states, $\Omega$. This is associated with $S$, Boltzmann's entropy as a measure of
Shannon information content of the system.

Then entropy increase, and also associated H-function, can be expressed for a lattice dynamics,

{\bf Conjecture 1} Occupancy of $N$ items on $M$ discrete states, $M>N$, evolving with dynamical
rules $\mathscr{D}$ necessarily increases $\Omega$, compare to the number of sampling if it were $M=N+1$.

This conjecture implies a diffusion process where by collection of particles moves into direction of empty portion of the space,
even rules goveringin the dynamics,  $\mathscr{D}$, is simple, given that it is physically plausable. Furthermore,
such conjecture also implies a arrow-of-time, naturally a definition follows,  \cite{kittel04}.

{\bf Definition 2}: Time's arrow is identified with change in entropy of material systems, that $\delta S \ge 0$.

In ideal setting the reversibility implies running dynamical rules backward would yield to an initial condition
again, hence a paradox. We address this by introducing counterfactual interventions.

\subsection{{do}-calculus interventional ensemble}

An analogy to $do-calculus$, a counterfactual dynamics on reversible dynamical evolution is induced by choosing a secondary ensemble.
This secondary ensemble, so called an interventional ensemble is introduced. A causal effect of sampling this secondary ensemble
on the same dynamical rules is stated.

{\bf Conjecture 2} ({\it H-Theorem do-conjecture}): Boltzmann's H-function provides a basis for entropy increase, it is associated with conditional probability of a system $\mathscr{S}$ being in state $X$ on ensemble $\mathscr{E}$. Hence, $P(X|\mathscr{E})$. Then, an irreversible evolution from time-reversal dynamics should use interventional notation, $P(X|do(\mathscr{E}))$. The information on how time reversal dynamics leads to time's arrow encoded on, how dynamics provides an interventional ensembles, $do(\mathscr{E})$.  The difference between evolution of set size of $|\Omega|$ and $|do(\Omega)|$,$\Delta H$, measures a causal effect, hence, an irreversibility.

\begin{figure}
  \includegraphics[width=0.5\textwidth]{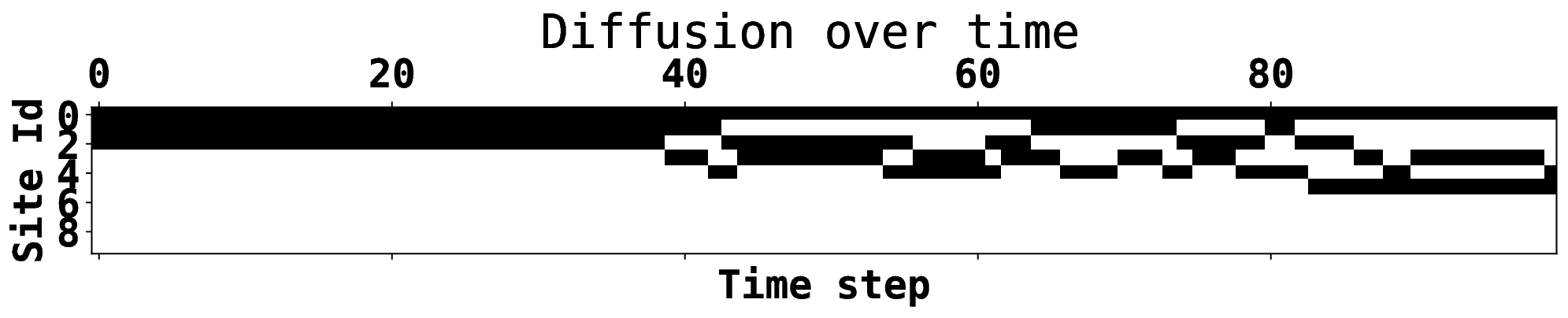}
  \caption{ {Ising-Conway Entropy Game} for M=10 sites with N=3 occupancy over time, for non-interventional dynamics.}
\end{figure}

\section{Random walkers for deterministic diffusion}

A system that mimics deterministic diffusion is introduced in demonstrating the conjectures. The system
uses idea of occupancy on discrete states via simple dynamical rules on a one dimensional lattice. $M$ sites, 1-dimensional
representation having $N$ occupancy, i.e., mimic a particle, We count  $\Omega$ as $k$ states with a count value of $|\Omega|=2^{k}$: boundaries
between two outermost occupied sites at a given time. This computation of $\Omega$ follows Conjecture 2. Mathematically, given configuration
on $M$ sites $C(t)$ at time $t$, whereas each component would take a value $c_{i}(t) \in \{0,1\}$. Hence, $k$ can be computed over time,
$$ k(t) = argmax \mathbb{I} \left[ C(t) \right] - argmin \mathbb{I} \left[ C(t) \right] $$
where $\mathbb{I}$ return the indices of $N$ $1s$ from $M$ sites . We consider this value as a proxy to H-function as well
as entropy. Initial condition is choosen to be fixed $N$ sites occupy a corner portion of the lattice as a
first step. This mimics increase in entropy, as diffusion progresses and $\Omega$ increases.

Following dynamic rules $\mathscr{D}$ are applied for $\mathscr{E}$, mimicking single-spin-flip dynamics \cite{suezen14}
and Conway's game of life \cite{conway}, i.e. {\it Ising-Conway Entropy Game} : At each time step we move a single occupied site
randomly or stays still, avoiding collision to neighbors akin to Pauli exclusion and site boundaries. We see a typical evolution of moves over
time for 10 sites and 3 occupancy in Figure 1, particles diffusing from the corner to empty space. In the case of
 $do(\mathscr{E})$ we move two occupied sites simultaneously, i.e., {\it dual-spin-flip dynamics}, obeying the
same dynamical rules.

\begin{figure}
  \includegraphics[width=0.4\textwidth]{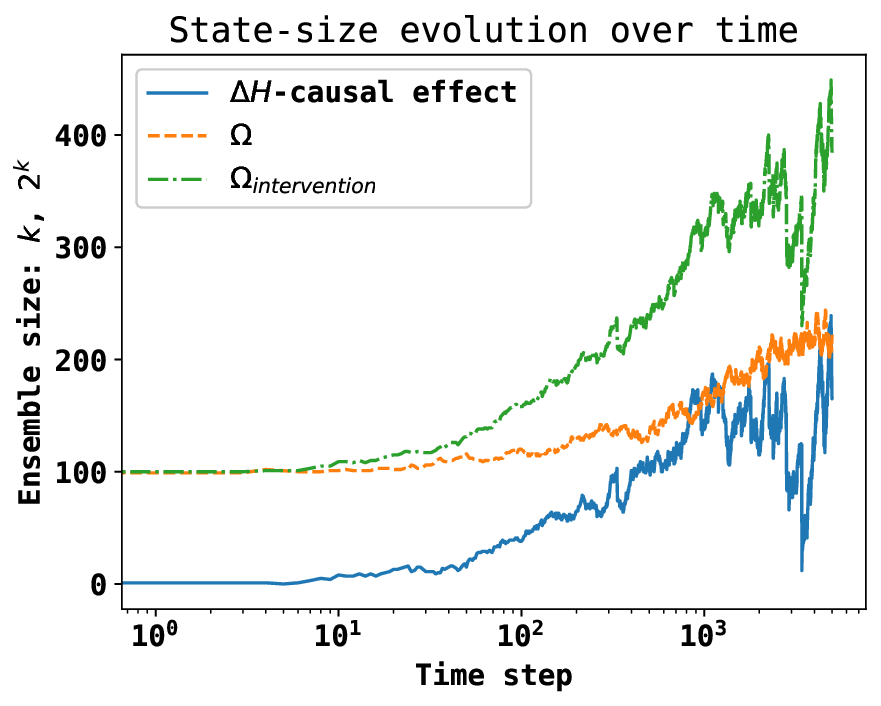}
  \caption{Evolution of $\mathscr{E}$ and $do(\mathscr{E})$ and causal effect over time for 1000 sites with 100 occupiers.}
\end{figure}

Following Conjecture 2 we compute both  size of $|\Omega|$ and $|do(\Omega)|$,$\Delta H$ for different $M$ and $N$ values,
that are large enough against size-effects.  The resulting deterministic diffusion, given sequence of random moves, shown in Figure 2, 3 and 4. We
observe that proposed causal effect is non-zero during games dynamic phase that the occupying grids are evolving. Such behaviour implying time-asymmetry
given a causal direction between base and interventional ensemble. Such as minor change in generating ensemble in dynamical rule gives a significant
intervention, demonstrating Conjecture 2 numerically, establishing causal connection between {\it single-spin-flip} and {\it dual-spin-flip} dynamics.

\begin{figure}
  \includegraphics[width=0.4\textwidth]{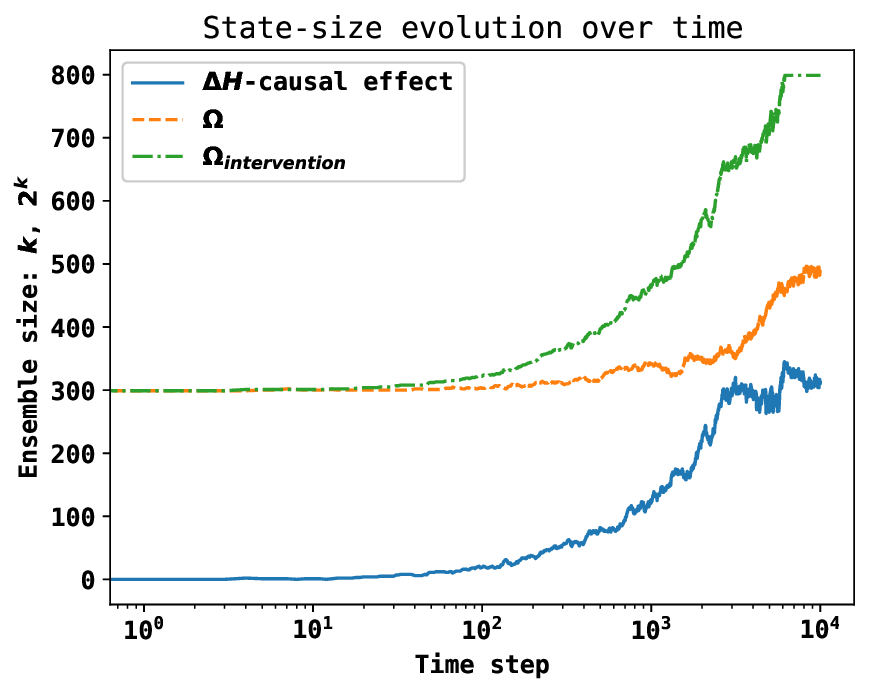}
  \caption{Evolution of $\mathscr{E}$ and $do(\mathscr{E})$ and causal effect over time for 800 sites with 300 occupiers.}
\end{figure}

\begin{figure}
  \includegraphics[width=0.4\textwidth]{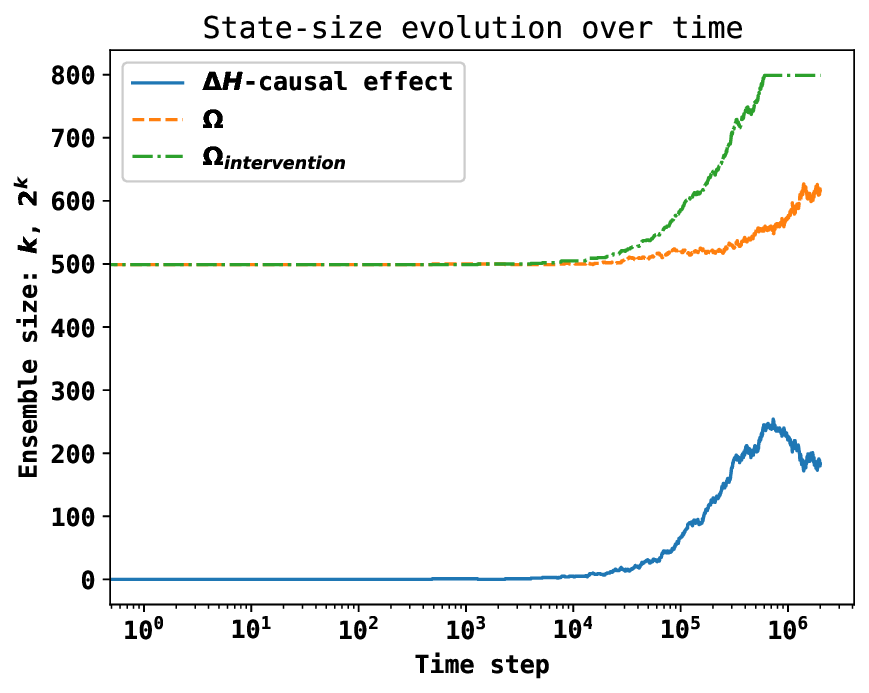}
  \caption{Evolution of $\mathscr{E}$ and $do(\mathscr{E})$ and causal effect over time for 800 sites with 500 occupiers.}
\end{figure}

\section{Conclusions}

Introducing an interventional ensemble resolves Loschmidt's objection, as $\Delta H$ is associated with conditional
probabilities of a system being in a given state over-time: A counterfactual intervention that is even a single
simple change induces assymmetric mechanistic histories in a deterministic fashion with a causal direction.
In other words, a relaxation time of two sampling schemes with slow and faster convergence on the identical dynamical
rules with a slight difference, dual vs. single flips, are genereted seperately. Practical implication of this finding for
simulating classical multi-body system lead to a requirement of introducing additional interventional ensemble
sampling schemes that search for a causal effect between two physical ensembles in the simulation, in measuring
physical properties obeying physical equation of motions, such as Boltzmann's Equation. This view is consistent
with {\it Molecular Chaos (Stosszahlansatz)} establishing a causal direction in time from a reversible microscopic
dynamics in a comparative setting as introduced here via {\it interventional ensembles}.

\nocite{*}
\bibliography{doh}

\end{document}